# Liquid metal intercalation of epitaxial graphene: large-area gallenene layer fabrication through gallium self-propagation at ambient conditions

S. Wundrack[1†], D. Momeni Pakdehi[1], W. Dempwolf[3,4], N. Schmidt[2,3], K. Pierz[1], L. Michaliszyn[1], H. Spende[2,3], A. Schmidt[2,3], H. W. Schumacher[1], R. Stosch[1], A. Bakin[2,3†]

[1]Physikalisch-Technische Bundesanstalt, Bundesallee 100, 38116 Braunschweig, Germany

[2]Institut für Halbleitertechnik, Technische Universität Braunschweig, Hans-Sommer Straße 66, D-38106 Braunschweig, Germany

[3]Laboratory of Emerging Nanometrology (LENA) der Technischen Universität Braunschweig, Langer Kamp 6 a/b, 38106 Braunschweig, Germany

[4]Institut für Technische Chemie, Technische Universität Braunschweig, Hagenring 30, 38106 Braunschweig, Germany

**Abstract.** We demonstrate the fabrication of an ultrathin gallium film, also known as gallenene, beneath epitaxial graphene on 6H-SiC under ambient conditions triggered by liquid gallium intercalation. Gallenene has been fabricated using the liquid metal intercalation, achieving lateral intercalation and diffusion of Ga atoms at room temperature on square centimeter areas limited only by the graphene samples' size. The stepwise self-propagation of the gallenene film below the epitaxial graphene surface on the macroscopic scale was observed by optical microscopy shortly after the initial processing without further physical or chemical treatment. Directional Ga diffusion of gallenene occurs on SiC terraces since the terrace steps form an energetic barrier (Ehrlich-Schwoebel barrier), retarding the gallenene propagation. The subsequent conversion of the epitaxial graphene into quasi free-standing bilayer graphene (QFBLG) and the graphene-gallenene heterostack interactions have been analyzed by XPS and Raman measurements. The results reveal a novel approach for controlled fabrication of wafer-scale gallenene as well as for two-dimensional heterostructures and stacks based on the interaction between liquid metal and epitaxial graphene.

## Introduction

The fabrication of large-area high-quality graphene samples has opened new fields of research to investigate the feasibility of creating new two-dimensional crystal structures. This applies in particular to semiconductor materials as well as to the fabrication of van der Waals heterostructures exploring their physical interactions at the nanoscale of condensed matter[1–9]. Ever since the demonstration of graphene was first achieved in 2004[10], a whole family of two-dimensional materials has been discovered and fabricated using the straightforward exfoliation technique[11–16]. For instance, hexagonal boron nitride (hBN) or transition-metal dichalcogenides (TMDC) such as tungsten disulfide ($WS_2$) and molybdenum disulfide ($MoS_2$) can be easily obtained by exfoliation from their bulk single crystals based on breaking the weak interlayer bonding similar to that of graphite[14–16]. The fabrication of large areas of the respective material is virtually impossible with this procedure, which allows the preparation of only crystal flakes with lateral dimensions of a few microns. Among 2D materials, elemental atomic layers of group III to group VI elements are gaining rapidly growing attention. Those classes of materials, so-called Xenes (silicene, phosphorene, stanene, etc.), cover a broad spectrum of physical properties from metallic to semiconducting and provide a wide range of potential applications in the areas of photonics, electronics, and energy conversion. Just recently, Kochat et al. demonstrated the fabrication of metallic gallenene[17]. At present, however, fabrication approaches, as well as properties investigations of gallenene, are at an early stage of research. It is already proposed that gallenene could be of particular interest for energy conversion applications[18]. Gallenane, the hydrogenated type of gallenene, is a promising material as a 2D flexible and stable metal for nanotechnological applications[19] and could be an extremely promising candidate for room temperature spintronics[20]. Even merely successful development of room temperature spintronics on the base of gallenane would be a huge breakthrough promoting the whole market of spintronic devices. Moreover, Ga polymorphism can be well implemented for the fabrication of phase-change plasmonic systems. The potential of Ga for pressure-driven phase-change plasmonic devices and plasmonic high-pressure surface-enhanced Raman spectroscopy (SERS) have been recently shown[21]. Using metals with a low melting point, such as elemental gallium or tin, could fundamentally change the exfoliation deposition technique due to their physical and chemical properties. The so-called solid-melt exfoliation might be a suitable approach to fabricate 2D materials in sizes up to the wafer-scale, as demonstrated in the previous studies[17,22,23]. Gallium (Ga) is a metallic element with an unusually low melting point of 29.7 °C[24,25] as compared to other metals. Strong electronic bonding states exist between two Ga atoms forming a Ga pair, whereas the bonding energy between Ga pairs is significantly lower[26]. Besides the liquid phase of Ga, several different solid-state modifications of Ga exist and are referred to as $α$- to $γ$-Ga phase occurring at defined pressure-temperature conditions[27] as well as in metastable supercooling phases[28,29]. The investigation of atomically thin Ga films on semiconducting substrates and their surface reconstruction behavior on these surfaces is still a matter of scientific studies[30]. The discovery of ultra-thin gallium oxide sheets, as well as the monochalcogenide gallium sulfide (GaS), has recently attracted attention since it provides a novel approach of the exfoliation technique utilizing the liquid instead of a solid phase of a metal[22]. Carey et al. have exfoliated atomically thin layers of gallium oxide from the surface of the liquid phase of Ga onto an oxidized silicon wafer, which was previously hydrophobized with a silane to create a patterned gallium oxide layer. Using surface chemistry, they converted the oxide layer into two-dimensional GaS[22]. Recently, Kochat et al. have demonstrated the exfoliation of "gallenene" – an atomically thin sheet of Ga – from the liquid Ga phase onto different substrates using the solid-melt exfoliation technique[17]. In this context, they demonstrated the extraction of the (010) and (100) crystal planes (zigzag and honeycomb-like structure) from $α$-Ga, which were stabilized at ambient conditions due to strong interactions with the underlying substrates as underpinned by density functional theory (DFT) calculations. The honeycomb phase of gallenene occurs even in an epitaxially grown monocrystalline bilayer, which is stabilized on the (0001) plane of gallium nitride (GaN)[31]. The gallenene layer has a thickness of only 0.55 nm and exhibits superconductivity at a higher transition temperature as compared to bulk Ga due to polarization effects from the underlying GaN substrate. Furthermore, the epitaxial growth of gallenene on a Si(111) surface has been demonstrated by Tao et al[30]. Also, the conversion of Ga into 2D GaN semiconductor layers by the chemical process of ammonolysis is of technological interest, as described by Al Balushi et al., was shown[32]. There, a thin Ga layer was prepared by incorporation of Ga atoms between an epitaxial graphene sheet and the SiC substrate at high temperature employing the decomposition of trimethylgallium as the Ga source. Very Recently, Briggs et al. have presented the intercalation of Ga, indium (In), and tin (Sn) atoms under epitaxial graphene[33]. The intercalation was carried out under vacuum conditions (300 Torr) using the evaporation of metals at high temperatures (700 – 800°C). The diffusion of Ga atoms in the proposed confinement heteroepitaxy (CHet) approach was activated by high defect densities in pre-treated epitaxial graphene, which acts as entrance slots for Ga atoms. Furthermore, the remaining Si dangling bonds on the SiC substrate promoted the detachment of the graphene buffer layer and conversion of epitaxial graphene into quasi-freestanding bilayer graphene (QFBLG).

† Correspondence and requests for materials should be addressed to S. W. (email: stefan.wundrack@ptb.de) or A. B. (email: a.bakin@tu-braunschweig.de)

The preliminary work of different research groups has shown a broad versatility of liquid Ga by downsizing the bulk material to the two-dimensional level. Based on our liquid metal intercalation technique (Li.M.I.T.) initially reported elsewhere[34], we demonstrate here an extremely simplified and a far more extensive large-scale deposition of gallenene at ambient conditions. The proposed technique enables the self-propagation and diffusion of Ga atoms through the intercalation of epitaxial graphene at room temperature. We show that neither high temperature and vacuum conditions nor the use of pre-treated graphene in oxygen plasma is necessary for successful metal intercalation. The Ga intercalation takes place through micropipes of the 6H-SiC substrate. The self-propagation and diffusion process of Ga atoms occurs along the terraces of the 6H-SiC substrate and affects the entire sample. Several analytical methods were used to investigate the physical properties of gallenene: optical microscopy, micro-reflection spectroscopy, scanning electron microscopy (SEM), X-ray photoelectron spectroscopy (XPS) and atomic force microscopy (AFM), as well as Laser Ablation Inductively Coupled Plasma Mass Spectrometry (LA-ICP-MS). In addition, we used confocal micro-Raman spectroscopy to evaluate the interlayer interactions between gallenene and graphene as well as van-der-Pauw measurements to assess their electrical properties.

## Results and discussion

**Deposition of gallenene on epitaxial graphene substrate.** A schematic representation of the fabrication process is depicted in Fig. 1a, including the observed self-propagation of gallenene. Optical images of large gallenene-covered areas as well as from regions acquired shortly after the liquid metal intercalation are shown in Fig. 1b, and c. In general, the deposition of a sizeable gallenene layer has been achieved by depositing a droplet of liquid Ga ($V \approx 10$ μl) onto the surface center of the SiC substrate (Fig. 1a, I) which is covered with epitaxial graphene on top of the (0001) lattice plane of 6H-SiC. The latter consists of monolayer graphene and the buffer layer, which is covalently bonded to the Si-terminated side of the SiC substrate[35–37]. The liquid metal intercalation was initiated by placing the epitaxial graphene sample with a Ga droplet on its surface onto an object slide and heating them to 120 °C. This treatment results in a temperature gradient from the annealing source to the sample surface so that a lower surface temperature of the epitaxial graphene sample can be assumed. However, since a higher process temperature could cause structural lattice defects in the graphene, the mentioned process temperature has not been exceeded. The Ga droplet was spread only to the lower side of the sample by wiping the liquid metal across the substrate surface using a squeegee, as indicated in Fig. 1a (I). The graphene layer, as well as the graphene buffer layer, withstands this treatment and has neither been removed nor damaged throughout the liquid metal intercalation, as examined by Raman spectroscopy. Hence, we suggest that liquid Ga acts as a lubricant during the process, decreasing the frictional force between the squeegee and the substrate surface. Figure 1b (I-III) represents three stages of a typical time-depending self-propagation of gallenene, which was recorded shortly after the spreading of the liquid Ga droplet and proceeded at room temperature. The color contrast in Fig. 1b (I-III) reveals large bright and dark areas across the substrate. Examining these regions by LA-ICP-MS measurements proves the existence of Ga unambiguously within the bright regions from its isotopic fingerprint (as shown later), while the dark areas correspond to uncovered epitaxial graphene is grown on 6H-SiC. A stepwise broadening of the bright colored areas can easily be observed using the red auxiliary line as the starting point. The propagation velocity of gallenene across the sample surface varies over time. Figure 1b (I-II) reveals a low propagation within the first 20 min including processing

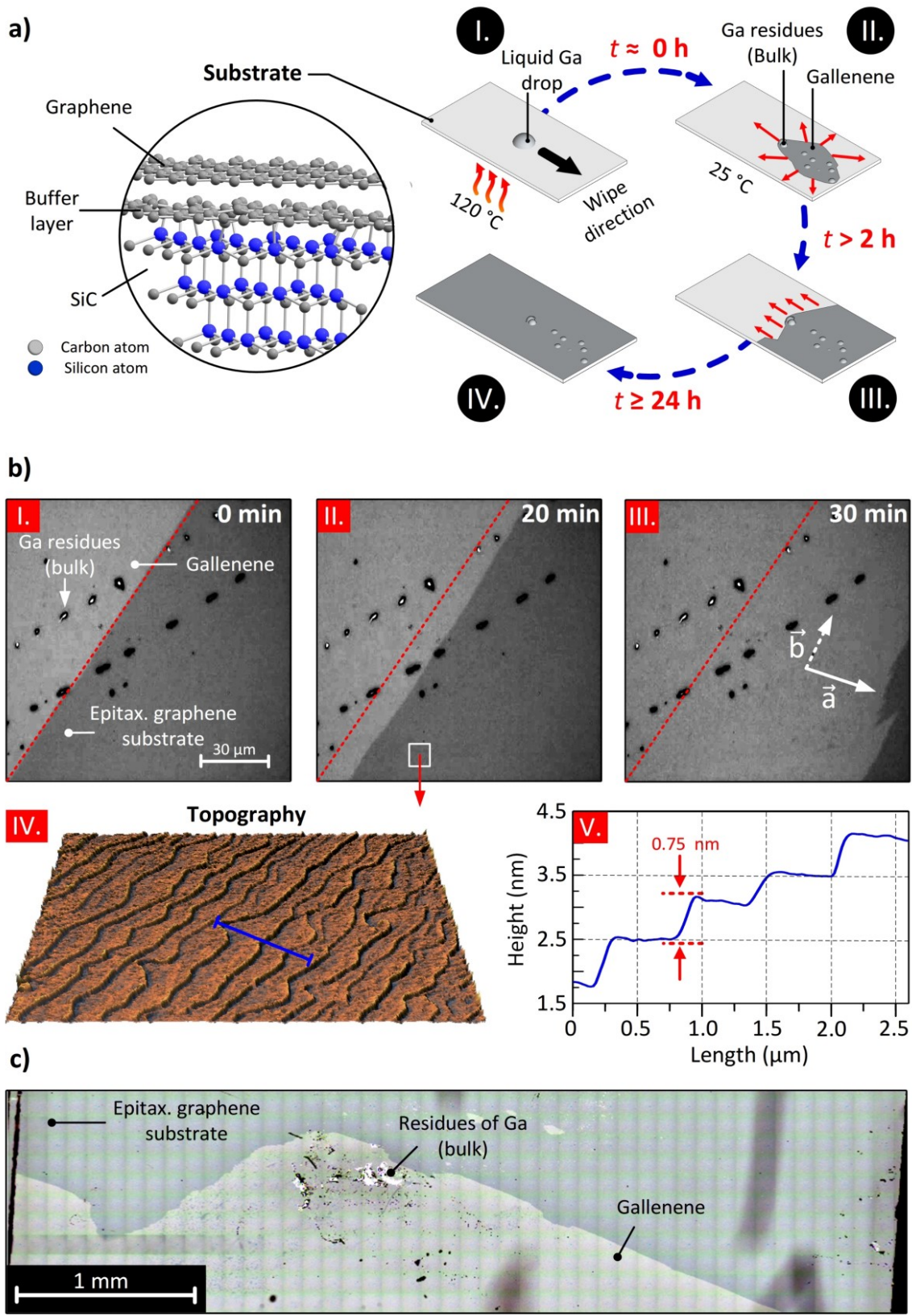

**Fig. 1 a)** Schematical representation of gallenene self-propgation caused by liquid metal intercalation technique (Li.M.I.T.). I. Initiation of the process by liquid Ga deposition and wiping the metal across the substrate. The inset shows schematically the sample structure consisting of monolayer graphene sheet on top of a graphene buffer layer which is covalently bonded to the underlying SiC substrate., II. Self-propgation of gallenene shortly after processing, III. Self-propagation of gallenene after 2 hours and IV. Gallenene completely intercalating the epitaxial graphene sample after >24 hours **b)** I-III. time-dependent self-propagation of gallenene. The red dashed line is used as start marker. IV. AFM topography of pure epitaxial graphene surface. V. Line profile of the substrate surface extracted from AFM topography. **c)** Optical micrograph of the gallenene-covered epitaxial graphene.

on the hot plate and cooling down to room temperature, followed by a significant increase in the subsequent period (Fig. 1b, II-III), which resulted in complete coverage of the investigated area within the next 10 min (Fig. 1b, III). The corresponding time-lapse video of the gallenene self-propagation at room-temperature, as shown in Fig. 1b is available in the supplementary information. Figure 1b contains three images extracted from the video showing different states of the self-propagation of gallenene that occurred fitfully in a non-continuous motion. The video, as well as Fig. 1b (III) reveals the existence of two different propagation directions indicated by the vector $\vec{a}$, which is approximately aligned to (1100) lattice plane of the SiC substrate and the vector $\vec{b}$ being aligned parallel to the terrace steps of SiC as revealed by AFM measurements (Fig. 1b, IV). The propagation velocity along $\vec{b}$ is significantly larger as compared to the direction of the vector $\vec{a}$. A detailed AFM analysis of the uncovered graphene surface shows an orthogonal direction of terrace steps to $\vec{a}$ having an average step height of 0.75 nm (Fig. 1b, IV-V). We, therefore, suggest that the diffusion of Ga atoms of gallenene is energetically favored along the SiC terrace ($\vec{b}$), whereas the terrace steps form an energetic barrier (Ehrlich-Schwoebel barrier) retarding the gallenene propagation along $\vec{a}$. Figure 1c clearly demonstrates that gallenene covers almost the whole width of the sample substrate while residues of Ga remain at the lower sample side. Finally, the self-propagation of gallenene affects the entire sample surface keeping the upper side of the sample free from any bulk Ga residues on the graphene surface. The upper region of the sample thus allows further analytical investigations (e.g., SEM, XPS, or Raman spectroscopy) to be carried out and enables the post-processing of the gallenene-graphene heterostack for preparing further experiments. The propagation velocity of gallenene decreased with progressing time, and it took a few hours until the lower sample side was covered entirely (Fig. 1a, II-III). The total coverage of the top half, however, was completed only after a few days by storage of the sample at room temperature (Fig. 1a, III-IV). The initiation of self-propagation of gallenene on the SiC substrate is decisively influenced by temperature as a crucial parameter. So far, the implementation of this approach entirely at room temperature without heating at the initial stage has not led to fabricate gallenene films with such large dimensions.

**Topological and chemical characterization of gallenene.** The formation and composition of a thin Ga film as metallene have been analyzed employing LA-ICP-MS, SEM, and AFM measurements, as shown in Fig. 2. In analytical chemistry, LA-ICP-MS has been evolved to a standard technique that can be used to identify and quantify inorganic impurities, for instance, in semiconductor materials[38–41]. Moreover, LA-ICP-MS mappings enable the lateral composition of thin films to be analyzed. Here, LA-ICP-MS mappings were measured over an area of (80 x 220) μm² and with an increment as well as a laser spot size of 20 μm encompassing the sharp transition from gallenene (bright-colored area) to pure epitaxial graphene (dark-colored area) as shown in Fig. 2a (red-colored box) and 2b. For better visualization, the pixels in Fig. 2c were linearly interpolated. The corresponding mass spectrum shows the signals of the two stable isotopes $^{69}$Ga and $^{71}$Ga (Fig. 2c) with small fractions of these isotopes, even occurring on pure epitaxial graphene, probably due to surface contaminations (blue mass spectrum, Fig. 2c). However, an increase in the Ga signal by more than one order of magnitude appearing in brightly colored areas (red mass spectrum, Fig. 2c) undoubtedly demonstrates the presence of Ga on the substrate. The chemical bonding states of graphene and gallenene have been studied by XPS and their physical interactions with each other by XPS and Raman spectroscopy. Results from the latter technique will be discussed later. The upper XPS spectrum in figure 2d shows the superposition of carbon-carbon bond energies in epitaxial graphene, indicating the presence of graphene buffer layer, graphene (284.61 eV) and SiC substrate (283.72 eV) at C1s core level[42]. Here, the binding energy of the graphene buffer layer is decomposed into $S_1$ (284.95 eV) and $S_2$ (285.54 eV) components and mimic various chemical bonding states within the graphene buffer layer[35,42,43] which, in turn, belong to chemical bonds between the buffer layer and the Si-face of the substrate[35,42–44]. Upon Ga deposition, the two signals S1 and S2 vanish, leaving only the binding energies of graphene and the substrate. The same XPS features have

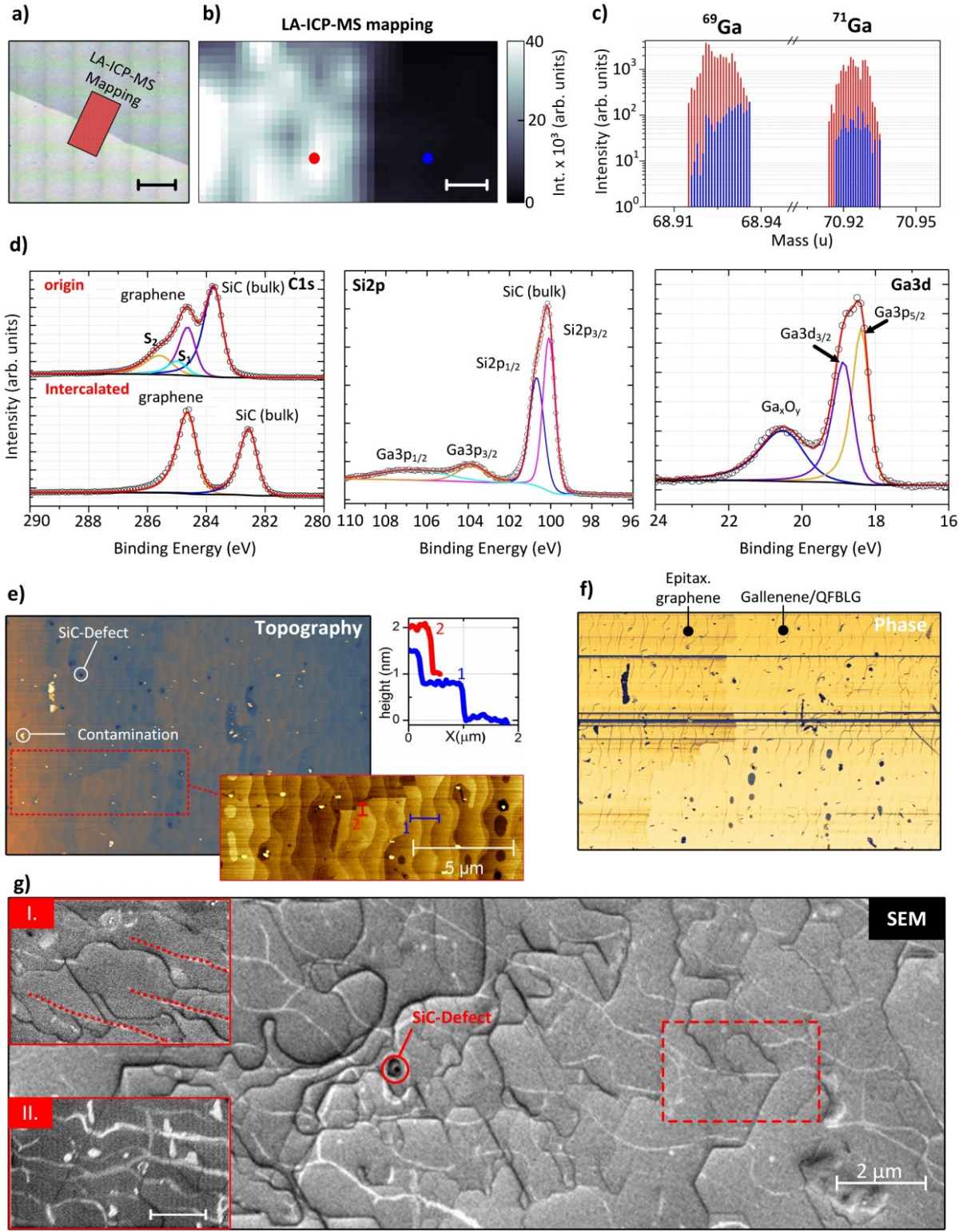

**Fig. 2 a)** Optical micrograph of the area considered for the LA-ICP-MS measurements. Scale bar corresponds to a width of 140 µm. **b)** False color mapping of gallenene on epitaxial graphene substrate using LA-ICP-MS. **c)** Mass spectrum of pure epitaxial graphene (blue) and of graphene covered with gallenene (red) showing signal intensities of gallium isotopes $^{69}$Ga and $^{71}$Ga. Both spectra result from areas from LA-ICP-MS mapping (blue and red circle). **d)** XPS spectra of the C1s and Si2p core level before and after Ga intercalation beneath epitaxial graphene. Ga3d core level indicating the presence of elemental Ga as well as Ga-oxide contents **e + f)** AFM topography and phase image of gallenene on epitaxial graphene. The inset shows a magnified representation of the measured topography (red dashed box) specifically the transition from epitaxial graphene substrate to gallenene covered areas. Line profile extracted from the topography (blue and red line) represents the thickness of gallenene (red) and the step height of the terraces (blue). **g)** scanning electron microscopy image (at 2 keV) of gallenene-intercalated area measured with in-beam mode SEM image. **g.I)** magnified SEM picture of area with gallenene. **g.II)** magnified SEM picture of pure epitaxial graphene substrate without gallenene. White stripes indicate terrace edges of SiC. Scale bar corresponds to a width of 2 µm.

been observed by Briggs et al.[33] as well as in the intercalation process of epitaxial monolayer graphene in different gas atmospheres resulting in a transformation into QFBLG[42]. In the latter case, at elevated temperatures, the gas molecules break the chemical bonds between the graphene buffer layer and the substrate and passivate at the same time the dangling bonds on the Si-face during the intercalation process[43–45]. As can be seen from XPS measurements, the Ga3d core level spectrum reveals the presence of Ga-Ga bonding states in gallenene at ~18.5 eV[46]. The metallic phase of gallenene is furthermore accompanied by small gallium oxide fractions at 20.49 eV[47]. Comparable oxide signals have been measured on top of the intercalated graphene bilayer by Briggs et al.[33], where a high-temperature approach in the vacuum has been employed. In contrast, we assume that small fractions of the gallium oxide are intercalated beneath epitaxial graphene since Li.M. I.T. was carried out at an ambient atmosphere. More importantly, such oxide signals were explicitly measured in areas that were free of Ga contamination of graphene upper surface resulting only from the liquid metal intercalation. Additional signals in the Si2p core level spectrum at ~103.8 eV and ~106.7 eV can be assigned to the Ga3p binding energies with spin $1/2$ and $3/2$ [46]. Using angel-resolved XPS measurements (ARXPS), attempts were made to detect possible evidence of bondings states of Si-Ga in the binding energy range near ~1116 eV of the Ga2p core level[48] as well as in the binding energy range around ~19 eV of the Ga3d core level spectrum[49]. Si-O bonding states at the Gallenene – SiC interface have been investigated in the Si2p core level spectrum at 103 eV[45,50,51]. The first state indicates binding energies between gallenene and the Si-face of the substrate, and the latter indicates a thin silicon oxide layer at the gallenene to substrate interface resulting from the presence of oxygen in oxidized gallenene (Fig. 2d, Ga3d core level). However, our measurements show no such bonding states between Si-Ga and Si-O, since these, if any, are below the detection limit of the XPS measurement. From that, we suggest that not only Ga is responsible for the intercalation process, but additionally, a small fraction of oxygen might also saturate some dangling bonds on the Si-face[50,51]. Therefore, both Ga and oxygen must be taken into account here. Nevertheless, XPS clearly shows the complete conversion of epitaxial graphene into QFBLG and thus underpins that the existing dangling bonds could be crucial for the intercalation, as has already been shown in the literature. We would like to point out that supercooled liquid Ga also has an anomalous physical behavior, which could be an essential criterion for intercalation at room temperature, which will be discussed later. The gallenene coverage across the substrate was also characterized by AFM. Figure 2e shows the topography of the SiC substrate, including the typical terrace structure, as already shown in the previous section on the pure epitaxial graphene surface. The topography image also shows a high density of small hole-like defects on the substrate, also known as micropipes. These surface defects occur in low crystal quality SiC wafers or could have been created during graphene fabrication employing the PASG process. The origin of the micropipes can be theoretically described by the Frank theory, according to which they often appear due to bunching of several screw dislocations to a super-screw dislocation[52]. In addition, the growth of epitaxial graphene stops at dislocations, and, under certain circumstances, these defects develop into patches of bilayer or even few-layer graphene[53]. The identification of the deposited gallenene sheets from AFM topography images is challenging because the measured gallenene topography is disturbed by the dominant SiC terraces. On the other hand, the AFM phase contrast image readily shows changes in the material properties, thus enabling the identification of pure epitaxial graphene and gallenene (Fig. 2f). A sharp transition boundary between areas of pure epitaxial graphene and large gallenene areas appears as a slight contrast change in the phase image (Fig. 2f). The difference in height between the pure substrate and the gallenene-covered areas is particularly evident, along with the terrace steps of the substrate (Fig. 2e). The average thickness of gallenene of about 1 nm can be observed in these areas (Fig. 2e, line no. 2, red line), which is approximately four times smaller compared to the value of solid-melt exfoliation approach of liquid Ga on Si(111)[17]. In contrast, the blue line shows an average terrace height of 6H-SiC of about 0.75 nm. In comparison to high-resolution TEM measurements on gallenene[33], our AFM results of gallenene thickness are in the range of 3 to 4 gallenene layers.

High-resolution In-Beam SEM images (Fig. 2g) has been used to reveal the topological difference between gallenene intercalated (Fig. 2g inset II.) and gallenene-free (Fig. 2g inset I.) areas of graphene on the SiC substrate. The presence of a gallenene leads to a brighter contrast compared to the pure epitaxial graphene region. The SEM-images also reveal the well-resolved characteristic terrace steps of the SiC wafer depicted as bright lines (white stripes) along the characteristic SiC terrace step edges, which remain visible even throughout the gallenene film (Fig. 2g I. noted in the inset using dashed red lines). Additionally, various dark lines appear within the gallenene film (Fig. 2g and inset I.). These lines clearly point out a symmetrical shape that could be forced by the hexagonal symmetry of the Si-face and indicate grain boundaries within the gallenene film, which is in good agreement with Raman measurements discussed below. It is worthwhile to mention that also an interesting reflectivity contrast observed in the SEM imaging of the SiC/Ga/BLG heterostack on the non-identical SiC terraces, which can be related to the stacking-order-induced doping valuation from the 6H-SiC stacking terminations[54].

**Investigation of interlayer interactions of graphene-gallenene.** Raman spectroscopy is an essential instrument for the characterization of 2D materials and is used here to investigate the lattice structure and interactions between graphene and gallenene. Raman spectra from pure epitaxial graphene (black) and from the gallenene - epitaxial graphene stack (red) are depicted in Fig. 3h. Both Raman spectra contain phonon bands that belong to the SiC substrate (marked with an asterisk). Characteristic phonon bands of graphene appear at 1580 and 2670 cm$^{-1}$, corresponding to the well-known G- and 2D peak. The physical origin of the D-, G- and 2D peak of graphene has been extensively investigated, and a description can be found elsewhere[55–59]. Creighton and Withnall have analyzed the space group symmetry of $α$-Ga, whereupon 4 Ga atoms occupy the primitive unit cell of a $D_{2h}^{18}$ space group and, thus, result in six Raman active phonon bands of $α$-Ga[26]. Their Raman spectroscopic experiments have shown that only one characteristic Raman band of solid $α$-Ga appears around 246 cm$^{-1}$, corresponding to the symmetrical stretching of Ga pairs. No such phonon band appeared in the Raman spectra of gallenene (Fig. 3e, red line), highly likely due to structural change of gallenene lattice compared to the $α$-Ga (bulk) resulting from the confinement at the graphene - 6H-SiC interface as observed by Briggs et al. using TEM[33]. There it is suggested that trilayer gallenene arises along the hexagonal lattice plane (111) of a distorted face-centered cubic (fcc) crystal symmetry of Ga(III) phase under high pressure between graphene and the hexagonal lattice plane (0001) of SiC substrate. Interestingly, Steenbergen et al. have calculated a new phase of trilayer gallenene structure using DFT calculations[60], which is comparable to the lattice structure of the beta phase of bulk Ga. The calculated trilayer structure occupies the same hexagonal lattice symmetry as the Si-face of SiC substrate and from our point of view indicating a low lattice mismatch between both materials. Especially the cross-sectional side view of the calculated trilayer gallenene crystal lattice resembles well with the cross-sectional TEM measurements given in the literature[33] and could, therefore, be an alternative model to the suggested Ga(III) phase[33]. The Raman spectrum of gallenene covered areas (Fig. 3h, red line) exhibits a strong signal between 45 and 750 cm$^{-1}$ with its maximum occurring at ~104 cm$^{-1}$ as compared to the Raman spectrum of the pure epitaxial graphene substrate (black line, Fig. 3h). However, the gallenene spectrum is not completely visible in the low-wavenumber range > 45 cm$^{-1}$ as these bands have been cut by an edge filter placed in the optical path of the spectrometer (Fig. 3h, red line). This spectral shape closely resembles that of a Boson peak, which is a typical effect arising in supercooled liquids[61], highly polycrystalline or amorphous materials[62]. Raman spectroscopic investigations of bulk Ga droplets have revealed a similar spectral line shape as compared to gallenene (Fig. S1, see supplementary) that was already observed by Lee and Kang even on thin $β$-Ga$_2$O$_3$ films[63]. In their work, the growth mechanism of $β$-Ga$_2$O$_3$ nanowires on

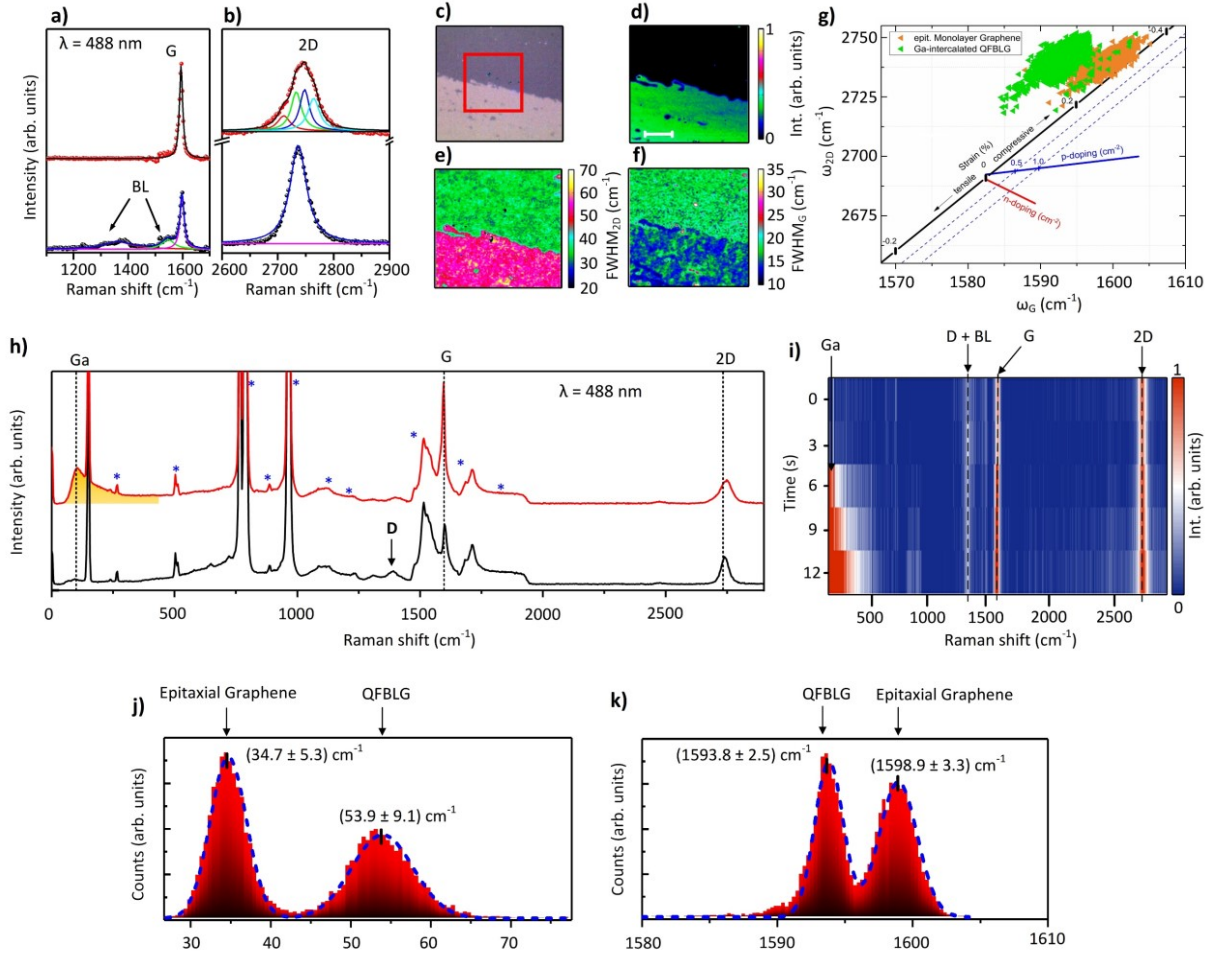

**Fig. 3 a+b)** Subtracted Raman spectrum of epitaxial graphene (blue line) and intercalated QFBLG (red line). **c)** Optical micrograph of the gallene-graphene boundary indicating the area selected for Raman mapping (red square). **d)** Measured Raman intensity signal of gallenene. Dark areas represent areas of pure epitaxial graphene substrate. Green-coloured areas correspond to gallenene **e)** Lateral 2D peak width distribution along the graphene-gallenene boundary. Green-coloured areas indicate regions of monolayer epitaxial graphene, whereas pink-coloured areas indicate the intercalation of epitaxial graphene caused by gallenene. Scale bar corresponds to a width of 5 μm **f)** G peak width distribution indicating electronic doping within graphene on gallenene-covered areas (blue-coloured areas). **g)** Strain-doping trajectory of epitaxial graphene (yellow dots) and intercalated QFBLG (green dots) **h)** Raman spectra of pure epitaxial graphene (black line) and gallenene-graphene heterostack (red line). Raman signal related to gallenene appears below 250 cm$^{-1}$ (dashed purple line). SiC phonon bands are marked with an asterisk. **i)** Time-dependent evolution of the phonon bands of graphene during the gallenene self-propagation. The intercalation of epitaxial graphene to QFBLG appears after 6 seconds. **j)** Histogram of 2D peak width of epitaxial graphene and intercalated QFBLG extracted from Raman mapping **k)** Histogram of G peak position width of epitaxial graphene and intercalated QFBLG extracted from Raman mapping.

monocrystalline crystal structure of $\beta$-Ga$_2$O$_3$, but they also observed two broadened bands at 115 and 703 cm$^{-1}$ on thin Ga films. Based on their Raman data, they suggested that a low crystallinity phase exists in those thin films, which cannot be assigned to known Raman active phonon modes of pure solid Ga or $\beta$-Ga$_2$O$_3$, but might instead indicate the presence of a Ga-rich oxide phase[63]. Hence, their data support our experimental results of Raman as well as XPS measurements, indicating a low-crystalline gallenene phase to be formed after the deposition onto the SiC substrate when using the solid-liquid exfoliation-intercalation technique. Furthermore, it is expected that supercooled Ga exhibits an anomalous liquid behavior at specific temperature and pressure conditions[64]. The anomaly manifests itself as density anomaly occurring as liquid phase separation, also called low- (LDL) and a high-density liquid (HDL), and as thermodynamic anomaly. Especially the change of liquid density of the supercooled Ga phase could support the intercalation process, in addition to the existing driving force of dangling bonds[33] which could act as a capillary force for Ga atoms. The change of liquid density of supercooled Ga might increase the mobility of Ga atoms, which thus enables a higher diffusion length of Ga atoms beneath epitaxial graphene, so a metastable liquid state of Ga arises under these conditions.
9

The Raman spectroscopic investigations of gallenene instead suggest a nanocrystalline phase or possibly even the presence of a metastable liquid phase of Ga at the microscale. A similar signal seems to appear in the Raman spectra of Kochat et al., showing the spectral fingerprint of pure gallenene and the gallenene-MoS$_2$ heterostack[17]. Unfortunately, a description of this signal is missing in their work. Overall, taking into account our measurements and the current literature, some questions about the gallenene lattice structure at the Si/graphene interface remain, so that further measurements have to be carried out for the clarification.

The lateral distribution of the individual graphene and gallenene phonon bands was measured using micro-confocal Raman mappings in combination with an automated non-linear fitting algorithm that extracts the characteristic parameters such as peak position and peak width (Fig. 3d, e, and f). All Raman spectra were corrected by subtracting a reference spectrum obtained from a pure 6H-SiC, since the second-order vibrational modes of 6H-SiC cover the Raman spectrum of graphene in a spectral range between 1200 cm$^{-1}$ and 2000 cm$^{-1}$ leading to a spectral superposition. Raman mapping is shown in Fig. 3d depicts the collected intensity distribution of gallenene across the epitaxial graphene. Black-colored areas reveal the absence of gallenene showing uncovered epitaxial graphene, whereas green-colored regions indicate the rise of the gallenene background signal in the Raman spectrum. The false-color image of the 2D peak width distribution in Fig. 3e illustrates a distinct contrast difference, representing a broadening of the 2D peak width on gallenene covered areas, which is consistent with the gallenene background signal distribution in Fig. 3d. A statistical evaluation of the 2D peak width of Fig. 3e underpins this effect by the formation of two separated Gaussian-like distributions (histogram, Fig. 3j) with the 2D peak widths being $(34.7 \pm 5.3)$ cm$^{-1}$ and $(53.9 \pm 9.1)$ cm$^{-1}$. The former distribution denotes 2D peak widths typically measured for epitaxially grown monolayer graphene[65], while the latter rather indicates bilayer graphene. Figure 3a and b depict the subtracted Raman spectra of pure epitaxial graphene (black line) as well as of the gallenene – graphene heterostack (red line). The Raman spectrum of epitaxial graphene (Fig. 3a and b) consists of the G and 2D peak related to monolayer graphene and is accompanied by flat and broadened phonon bands of the buffer layer in the Raman spectral range from 1200 cm$^{-1}$ to 1600 cm$^{-1}$ [66]. The graphene buffer layer background vanishes on gallenene-intercalated areas, as shown in Fig. 3h (red line) followed by an increase of the G peak intensity, which confirms the XPS results. Furthermore, the 2D peak line shape changes in gallenene intercalated areas, as depicted in Fig. 3b (red line) and can be fitted by four Lorentzian curves (for more details see supplementary Fig. S2 and S3). The characteristic low energy out-of-plane phonon modes (around 100 cm$^{-1}$ and 1800 cm$^{-1}$) of bilayer graphene[67] are superimposed by the background signal of gallenene and the second-order vibrational modes of SiC and were, therefore, not observed in our Raman measurements. The time-dependent evolution of the above-mentioned Raman spectral changes is represented in Fig. 3i showing the dynamical spectral change of the graphene Raman spectrum during the intercalation process. The intensity of the G and the 2D peak widen instantly as the Ga signal rises.

The electronic, as well as physical interactions between gallenene and graphene, were also evaluated using confocal micro-Raman spectroscopy[68–73]. Electronic doping of graphene significantly affects its Raman spectroscopic properties by changing especially the G peak position and peak width, typically resulting in stiffening and sharpening of the G peak[68,70]. Fig. 3f depicts the G peak width mapping of graphene, indicating predominantly blue and green-colored areas corresponding to G peak widths of 12 and 16 cm$^{-1}$. While the green areas indicate the absence of carrier doping in epitaxial graphene, the blue regions correspond to slight carrier doping of approximately $5 \times 10^{12}$ cm$^{-2}$ [70] introduced due to the interaction with gallenene, finally changing the carrier density of bilayer graphene. Furthermore, carrier doping occurs only on gallenene covered areas and is delimited by a sharp border, as shown in Fig. 3f. In addition, Fig. 3f depicts a non-homogenous distribution of the G peak width on gallenene covered areas, which implies rather heterogeneous doping across the graphene-gallenene heterostack. Due to this effect, Van der Pauw (VdP) measurements have been carried out to calculate the carrier density ($n$ or $p$) and carrier mobility ($\mu$) as a function of temperature in the sample[74]. The VdP measurement was performed without applying any lithography processing on the samples. The VdP measurement setup and sample preparation can be found in ref. [36]. The low carrier mobility of $\mu \approx (300 \pm 50)$ cm$^2$ V$^{-1}$ s$^{-1}$ as well as strong electron doping of $n \approx (4.5 \pm 0.1) \times 10^{12}$ cm$^{-2}$ and a sheet resistance $R_s \approx (4500 \pm 50)$ Ω have been measured on SiC - gallenene - QFBLG samples (SiC/Ga/QFBLG). This lower mobility is mainly due to the temperature-dependent electron-phonon scattering that prevails at higher temperatures. Thereby, $n \approx (1.3 \pm 0.1) \times 10^{12}$ cm$^{-2}$, $\mu \approx (1350 \pm 50)$ cm$^2$ V$^{-1}$ s$^{-1}$, and $R_s \approx (3700 \pm 50)$ Ω are calculated at cryogenic temperatures ($T = 4.5$ K). Here, the $n$-doping is a superposition of the carriers induced by both SiC and gallenene that make the free-standing top graphene bilayer $n$-type but not $p$-type, as would be the case, for example by hydrogen intercalation[36,75]. In addition, the SiC/Ga/QFBLG sample shows much lower mobility both at room temperature and at low temperatures compared to the QFBLG sample prepared using hydrogen intercalation in the same VdP regime, $\mu \approx 2700$ cm$^2$ V$^{-1}$ s$^{-1}$ ($T = 295$ K) and $\mu \approx 3350$ cm$^2$ V$^{-1}$ s$^{-1}$ ($T = 2.2$ K)[36], indicating a higher charge carrier scattering in the SiC/Ga/QFBLG system. Such strong electron doping concentration could be relevant for plasmonic applications such as substrates for surface-enhanced and tip-enhanced Raman spectroscopy (SERS and TERS) or tip-enhanced photoluminescence spectroscopy (TEPL). In this context, the optical properties of gallenene have been investigated using other optical techniques and are available in the supplemental information (Fig. S5), but are not the subject of discussion in this article.

The strain level within the graphene crystal lattice can be significantly affected by the intercalation process[45]. In this context, we have noticed a shift of the G peak position from $(1598.9 \pm 3.3)$ cm$^{-1}$ down to $(1593.8 \pm 2.5)$ cm$^{-1}$ (Fig. 3k), indicating a slight reduction of the strain level. Nevertheless, the G peak position is also superimposed by electronic doping[76], as has already been proved by the reduced G peak width. A more detailed analysis by using a strain-doping trajectory[72] for the G- and 2D-peak has shown that compressive strain in epitaxial graphene of $\varepsilon \sim 0.3$ is slightly decreased down to $\varepsilon \sim 0.2$ in QFBLG after the Ga intercalation. These experimental findings, however, cannot be confirmed by literature data, since an extensive study of the strain-doping trajectory for bilayer graphene is still missing. The trajectory is shown in Fig. 3g has been taken from previous work, as explained elsewhere[45]. The remaining compressive strain within graphene indicates strong interactions between gallenene and graphene. Considering the physical interaction between graphene and gallenene, the introduction of defects into the graphene lattice should also be considered. The investigated areas of our samples do not reveal any increased defect density of graphene after the gallenene spreading (Fig. S4, see supplementary), proving the advantage of using micropipes within the SiC substrate and corresponding holes in graphene for epitaxial graphene intercalation. Ga intercalation through highly defective epitaxial graphene, which was pre-treated by plasma etching, is followed by the "self-healing" of the graphene lattice, reducing the defect density. However, Raman spectra of Briggs et al. indicate that small defect densities remain after the self-healing process suggesting a slightly reduced crystal quality of QFBLG[33]. It can be concluded that defect density engineering of epitaxial graphene is a critical technology for creating high-quality heterostructures through intercalation. Al Balushi et al. demonstrated that graphene wrinkles enhance the nucleation process of Ga[32]. Other research groups have shown that the infiltration of alkali atoms can be achieved through wrinkles, which act as penetration sides[77]. In our case, the microholes present in both graphene and SiC act as nucleation centers for Ga atoms due to a decreased surface energy and support the intercalation of Ga atoms by enabling them to get under the buffer layer.

# Conclusions

In summary, we have demonstrated a novel approach to fabricate large-area gallenene layer through the intercalation liquid metal beneath epitaxial graphene at ambient conditions by self-propagation and diffusion of Ga atoms at room temperature. Liquid Ga penetrates through substrate defects of SiC, such as micropipes, and intercalates the graphene buffer layer, which finally converts the epitaxial graphene into QFBLG, as demonstrated by XPS and confocal micro-Raman spectroscopy. Furthermore, we suggest that the resulting gallenene film at the graphene-SiC interface has rather nanocrystalline properties and is in a metastable state, respectively. The self-organized process of Ga diffusion does neither require specific environmental conditions to be met nor any additional physical or chemical treatments. The diffusion of Ga atoms of gallenene is energetically favored along the SiC terrace, whereas the terrace steps form an energetic barrier (Ehrlich-Schwoebel barrier), retarding the gallenene propagation. We also propose that dangling bonds at the SiC – buffer layer interface trigger the intercalation process of Ga along the Si-face of 6H-SiC, which might also be supported by the liquid-liquid-phase transition of Ga at high pressure between the graphene/SiC interface. Symmetrical shapes within the gallenene film have been observed using SEM, which indicate domain borders or grain boundaries. The formation of such shapes could be forced by the hexagonal lattice symmetry (0001) of the Si-face of the 6H-SiC. The presented liquid metal intercalation technique (Li.M.I.T.), in combination with large-area epitaxial monolayer graphene sheets, could open a new route towards a controlled fabrication of homogenous large-area layers and 2D stacks employing liquid metals. The intercalated metallene layer (based on Ga, In, Sn, etc.) at the graphene-SiC interface could be converted into corresponding semiconductor derivatives such as an oxide or nitride layer, which would also lead to new 2D-stacked polytypes which do not exist under normal conditions.

# Materials and methods

**Substrate fabrication and treatment.** Epitaxial graphene was grown on the Si-face of SiC samples (5 x 10) mm² cut from a semi-insulating 6H-SiC wafer with a nominal miscut angle of about 0.06° towards ($1\bar{1}00$) surface. The graphene samples were prepared according to the polymer-assisted sublimation growth (PASG) technique, which involves polymer adsorbates to be formed on the SiC surface by liquid phase deposition from a solution of a photoresist (AZ5214E) in isopropanol followed by sonication and short rinsing with isopropanol. The graphene layer growth was processed at 1750°C (argon atmosphere ~1 bar, 6 min, zero argon flow) with a pre-vacuum-annealing at 900°C[35,36]. The applied PASG method allows the growth of large-area ultra-smooth, homogenous monolayer graphene, with almost an isotropic resistance characteristics[37], thus is very well suited as the basis for the fabrication of large-area two-dimensional gallenene sheets by solid-melt intercalation technique as described in this work. Commercial Ga (99.99 % purity) was purchased from Heraeus. More detailed information on the gallenene fabrication can be found in "Synthesis of gallenene on epitaxial graphene" within the section "Results and discussion."

**Measuring equipment.** Raman measurements of graphene and gallenene were acquired at ambient conditions with a Witec Alpha 300 RA (Witec GmbH) Raman spectrometer equipped with a frequency-doubled Nd:YAG laser emitting at 488 nm ($E_L$ = 2.54 eV) and a 100x (N.A. 0.9) objective to focus the excitation laser onto the sample surface. A LabRAM Aramis Raman spectrometer (Horiba) has been used in the case of excitation wavelengths of 532 nm ($E_L$ = 2.32 eV) and 633 nm ($E_L$ = 1.96 eV). Confocal micro-Raman mappings were recorded over (20 x 20) µm² scan areas in backscattering mode using a piezo-driven xy-stage (PI) and a scanning step size of 0.1 µm. AFM measurement was carried out using the Nanostation AFM (S.I.S), providing pronounced resolution for precisely resolving structures down to 0.25 nm as well as the AFM station of the Witec Alpha 300 RA RA (Witec GmbH). SEM investigations were carried out inside a Tescan Mira 3 GMH FE-SEM. A high surface sensitivity was reached using acceleration voltages between 1 and 5 kV and small beam currents. The in-lens SE detector offered sufficient SNR, high contrast, and high resolution for the small working distances used. Besides the topographic information mainly provided by the asymmetrically mounted SE detector, the additional in-lens SE detector also images the work function (e.g., material variations) of the sample. LA-ICP-MS mappings were measured by using a laser ablation system (NWR 213, New Wave Research Inc.), which is equipped with a frequency quintupled Nd:YAG deep UV laser emitting at 213 nm ($E_L$ = 5.82 eV) focusing the laser with a pulse energy of about 0.1 mJ and a spot diameter of 20 μm. Ablated materials were investigated by means of a double-focusing magnetic sector field high-resolution ICP mass spectrometer (Element XR, Thermo Scientific Inc.) analyzing the isotopic ratios of $^{69}$Ga and $^{71}$Ga. XPS measurements were obtained on an AXIS Supra system (Kratos Analytical Ltd.) using a monochromatized Al Kα X-ray source. Survey scans were acquired using a pass energy of $E$ = 160 eV (not shown). High-resolution core-level spectra (C1s, Si2p, and Ga3d) were obtained using a pass energy of $E$ = 20 eV. Data analysis was performed using CasaXPS software.

## Acknowledgments

S. Wundrack acknowledges support from the Braunschweig International Graduate School of Metrology B-IGSM. D. Momeni Pakdehi acknowledges support from the school for contacts in nanosystems (NTH). The work has received funding from the European Metrology Programme for Innovation and Research (EMPIR) under grant 17FUN08 co-financed by the Participating States and from the European Unions Horizon 2020 research and innovation program. The work was also co-funded by the Deutsche Forschungsgemeinschaft (DFG) under Germany's Excellence Strategy – EXC-2123 Quantum Frontiers (390837967).

# Supplementary: Liquid metal intercalation of epitaxial graphene: large-area gallenene layer fabrication through gallium self-propagation at ambient conditions

S. Wundrack[1†], D. Momeni Pakdehi[1], W. Dempwolf[3,4], N. Schmidt[2,3], K. Pierz[1], L. Michaliszyn[1], H. Spende[2,3], A. Schmidt[2,3], H. W. Schumacher[1], R. Stosch[1], A. Bakin[2,3†]

[1]Physikalisch-Technische Bundesanstalt, Bundesallee 100, 38116 Braunschweig, Germany

[2]Institut für Halbleitertechnik, Technische Universität Braunschweig, Hans-Sommer Straße 66, D-38106 Braunschweig, Germany

[3]Laboratory of Emerging Nanometrology (LENA) der Technischen Universität Braunschweig, Langer Kamp 6 a/b, 38106 Braunschweig, Germany

[4]Institut für Technische Chemie, Technische Universität Braunschweig, Hagenring 30, 38106 Braunschweig, Germany

This supplemental material presents:

1. Raman spectrum of Ga droplet (bulk)

2. Non-linear curve fitting of the 2D peak corresponding to intercalated graphene by gallenene

3. Phonon dispersion measurements of intercalated graphene by gallenene

4. Raman mapping of intercalated graphene and absence of lattice defects in graphen

5. Optical properties of gallenene-QFBLG heterostack using micro-reflectance and transmission spectroscopy

6. Video: Self-propagation of gallenene across epitaxial graphene sample

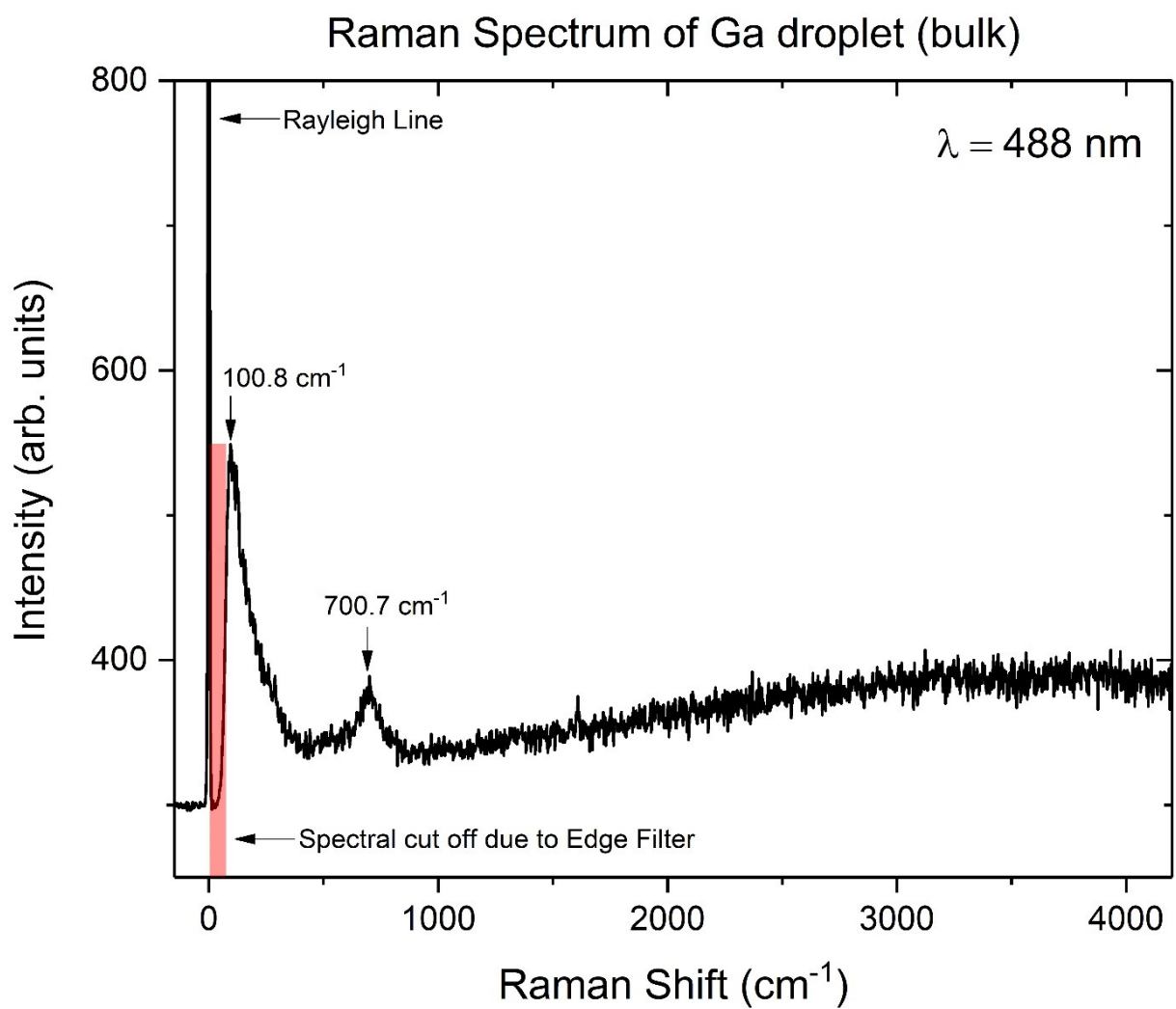

**Fig. S1** Raman spectrum of Ga droplet (bulk)

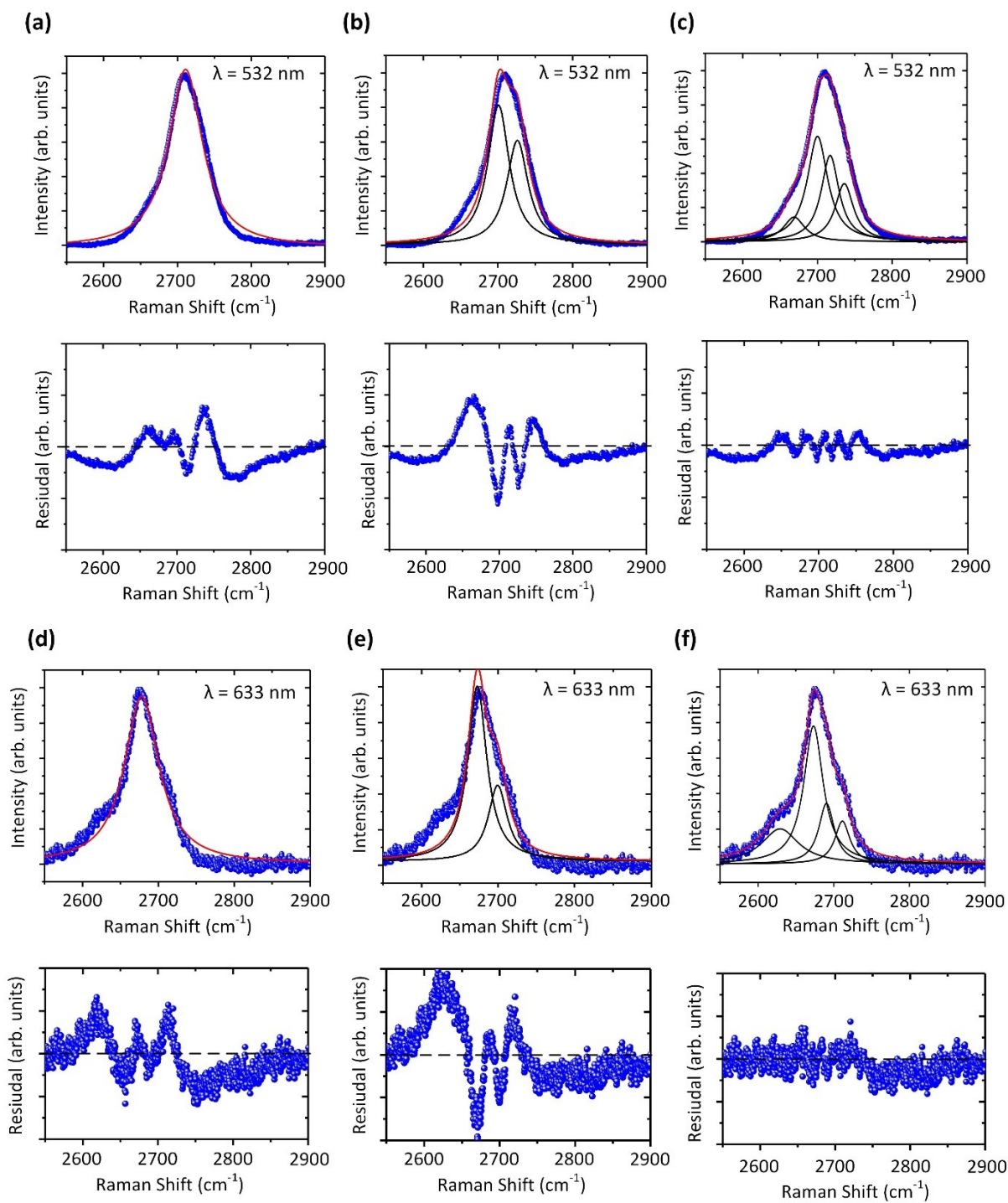

**Fig. S2** Non-linear curve fitting of the 2D peak corresponding to bilayer graphene after solid-melt exfoliation showing a satisfying curve fit by using 4 Lorentzian curves as proven by a clearly reduced residual distribution.

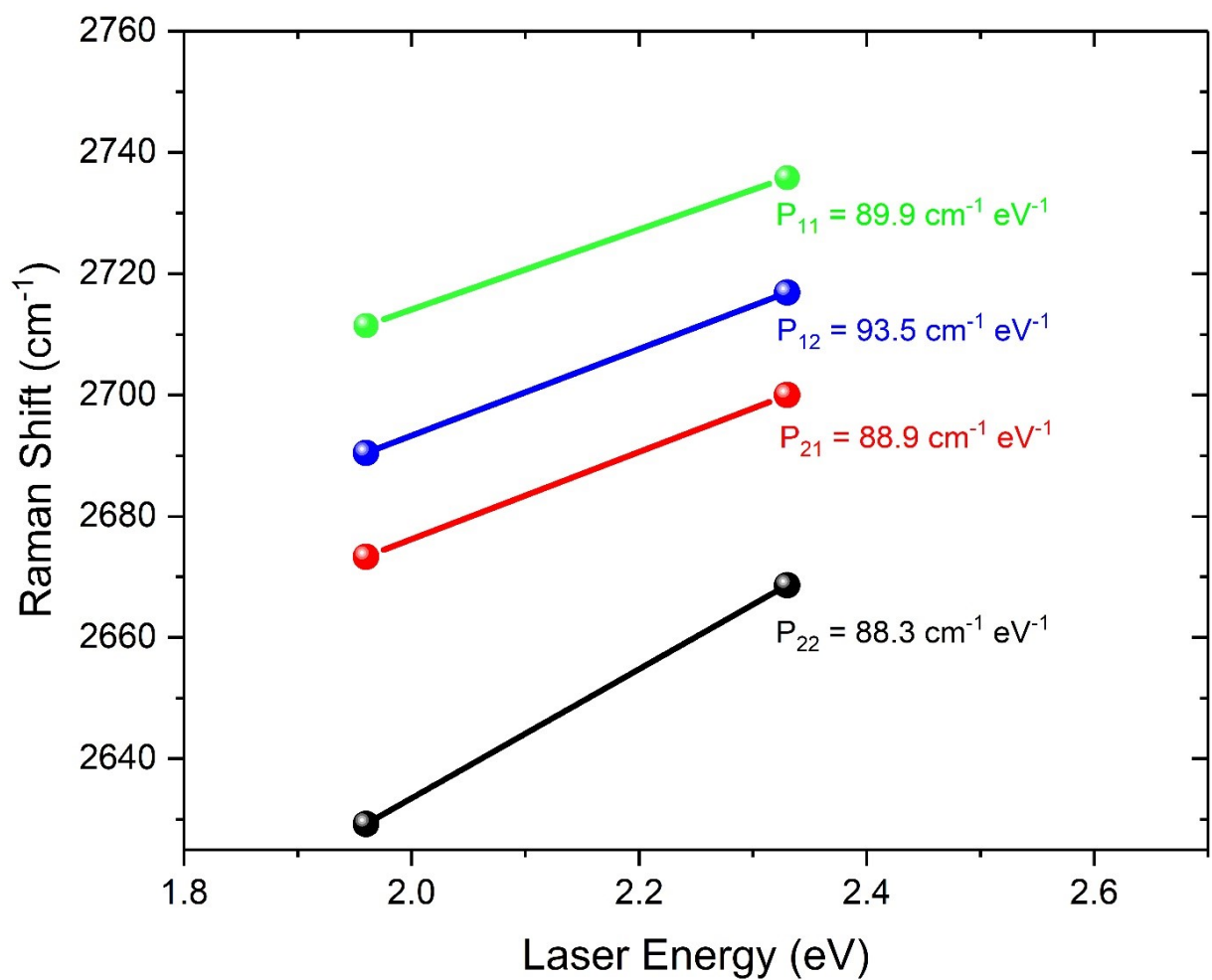

**Fig. S3** Phonon dispersion of the 2D peak at different Laser energies deconvoluted by 4 Lorentzian curves ($P_{11}$, $P_{12}$, $P_{21}$, $P_{22}$) from Fig. S2c and f

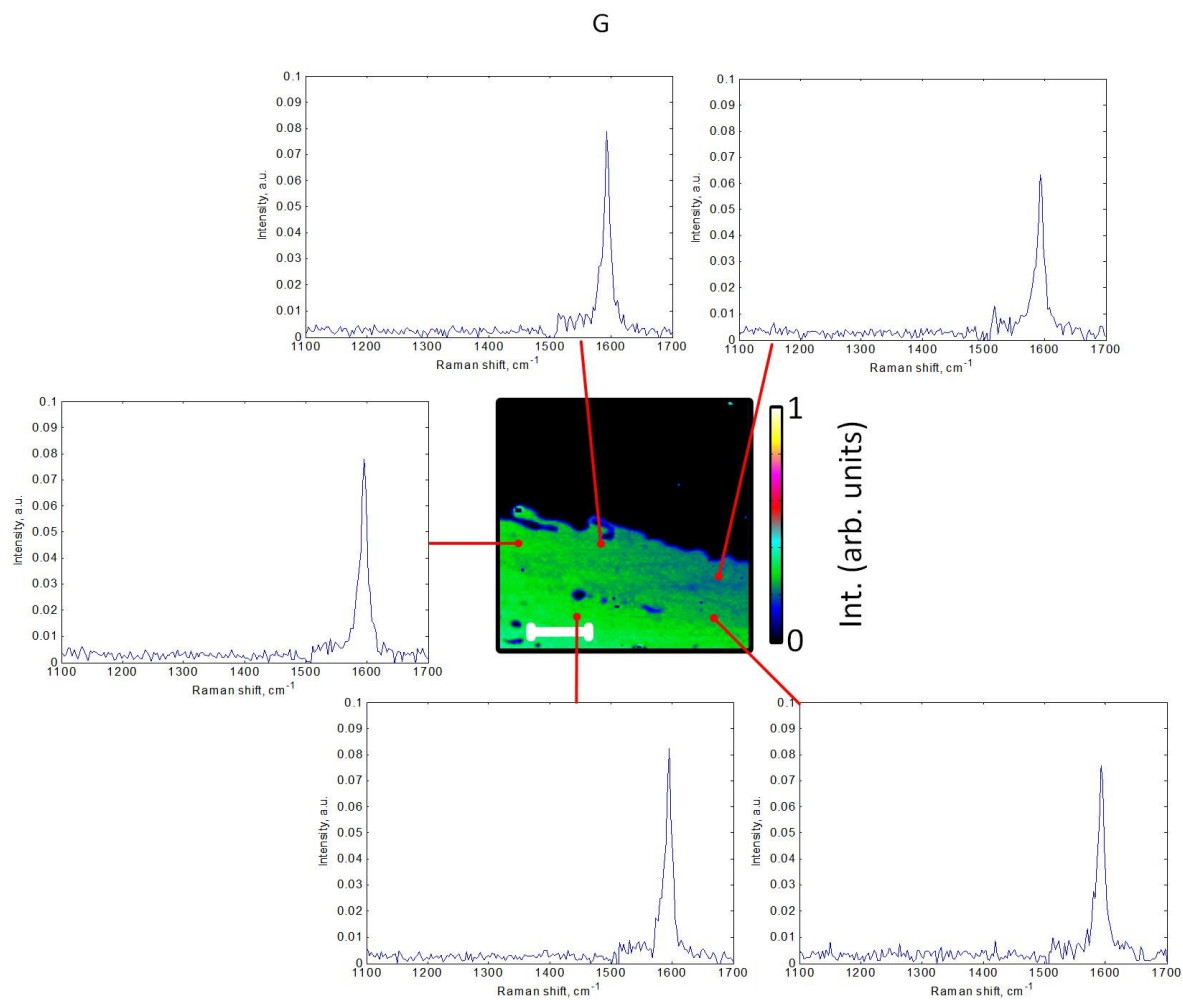

**Fig. S4** Raman mapping of epitaxial graphene sample after solid-melt exfoliation. False color mapping corresponds to the acquired Ga signal within the Raman measurements. Green colored areas indicate gallenene, whereat black colored areas prove its absence. Surrounding Raman spectra prove the absence of D peak (≈ 1340 cm$^{-1}$), evidencing no lattice defects in bilayer graphene in this investigated area after gallenene spreading as well as the removal of the buffer layer phonon bands.

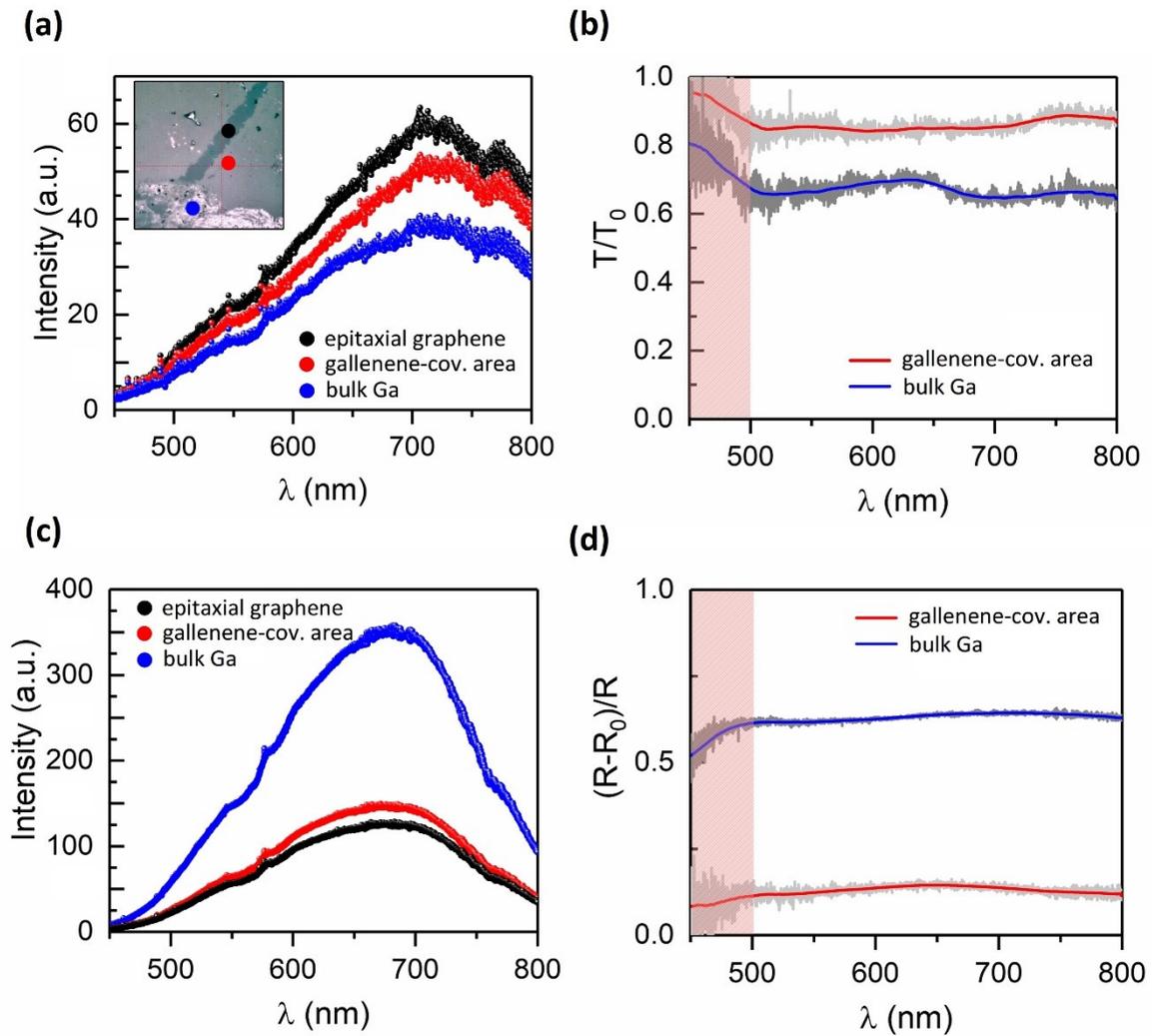

**Fig. S5** a) Reflectance intensity of pure epitaxial graphene (black), graphene areas covered with gallenene (red) and bulk Ga (blue). b) Differential reflectance spectrum of gallenene (red) and bulk Ga (blue) calculated according to Frisenda et al.(doi: 10.1088/1361-6463/aa5256). The grey-coloured line represents the calculated standard deviation. The red-coloured area (< 500 nm) marks spectral region of an increased uncertainty caused by the collapse of the spectral power distribution of the halogen lamp approaching the UV range.